\documentclass[prd,aps,preprint,tightenlines,superscriptaddress,nofootinbib,unsortedaddress]{revtex4-1}
\usepackage{amssymb}
\usepackage{amsmath}
\usepackage{epsfig}
\usepackage{subfig}

\begin{document}
\title{Universality of multi-particle production in QCD at high energies}

\author{Fabio Dominguez}
\affiliation{Institut de Physique Th\'{e}orique, CEA-Saclay, F-91191 Gif-sur-Yvette, France}
\author{Cyrille Marquet}
\affiliation{Physics department, Theory Unit, CERN, CH-1211 Geneva, Switzerland}
\affiliation{Departamento de F\'\i sica de Part\'\i culas and IGFAE,  Universidade de Santiago de Compostela, 15782 Santiago de Compostela, Spain}
\author{Anna M. Stasto}
\affiliation{Department of Physics, Pennsylvania State University, University Park, PA
16802, USA}
\affiliation{RIKEN BNL Research Center, Building 510A, Brookhaven National Laboratory,
Upton, NY 11973, USA}
\affiliation{H. Niewodnicza\'nski Institute of Nuclear Physics, Polish Academy of Sciences, Krak\'ow, Poland}
\author{Bo-Wen Xiao}
\affiliation{Institute of Particle Physics, Central China Normal University, Wuhan 430079, China}
\affiliation{Department of Physics, Pennsylvania State University, University Park, PA 16802, USA\\}

\begin{abstract}
\vspace{0.5cm}

By studying the color structure of multi-particle production processes in p+A-type (dilute-dense) collisions,
we find that higher-point functions beyond typical dipoles and quadrupoles, e.g., sextupoles, octupoles, etc.,
naturally appear in the cross sections, but are explicitly suppressed in the large-$N_c$ limit. We evaluate the sextupole in the McLerran-Venugopalan model and
find that, in general, its analytical form cannot be written as combination of dipoles and quadrupoles.
Within the Color
Glass Condensate framework, we present a proof that in the large-$N_c$ limit, all multi-particle production
processes in the collision of a dilute system off a dense  can, up to all orders in $\alpha_s$,
be described in terms of only dipoles and quadrupoles.
\end{abstract}
\maketitle

\section{Introduction}

Calculating cross sections for multi-particle (multi-jet)
production processes can be theoretically challenging when a
resummation of multiple interactions is needed, as is the case
in large parton density environments such as the small-$x$ regime accesible in high energy nuclear collisions. The main
complication comes from the fact that partons scatter coherently
and cannot be regarded separately. In particular, it is very important
to appropriately consider the color structure of the multi-particle states since color
conservation plays an important role as a source of correlations.

The standard way of calculating such multiple scattering processes
is to consider the small-$x$ gluons as an external field where high-energy
probes scatter in an eikonal way, see Ref.~\cite{Kovner:2001vi}. Under that framework, each
parton traversing the field contributes to the scattering
amplitude with a Wilson line in the appropriate representation
(fundamental for quarks, adjoint for gluons) at a fixed transverse
coordinate. Provided they are put together in the correct order,
the Wilson lines account for the color flow of the
process under consideration. After averaging (summing) over
initial (final) colors at the cross section level, one is left with a product of traces involving Wilson
lines in the fundamental representation and color matrices which
are contracted with adjoint Wilson lines. These adjoint Wilson
lines can be subsequently replaced by two fundamental Wilson lines
 by means of the identity
\begin{equation}
W^{ab}(x)=2\,\text{Tr}\left[t^aU(x)t^bU^\dagger(x)\right],
\end{equation}
where $W(x)$ stands for a Wilson line in the adjoint representation at a fixed transverse coordinate, $U(x)$ is the corresponding Wilson line in the fundamental representation, and $t^a$ are color matrices in the fundamental representation also. The color matrices can be removed using the Fierz identity
$t^a_{ij}t^a_{kl}=\frac{1}{2}\delta_{il}\delta_{jk}-\frac{1}{2N_c}\delta_{ij}\delta_{kl}$.
After these manipulations one is left with a sum of products of
traces of Wilson lines in the fundamental representation only.

In order to calculate measurable observables in this framework,
including cross sections for multi-particle production, it is
necessary to take a weighted average over the possible external
field configurations. This averaging process introduces
non-trivial correlations between the fields entering the different
Wilson lines, therefore giving rise to the afore mentioned
coherent scattering involving possibly all the partons appearing
in the process. The physics encoded in this field average is
inherently non-perturbative and therefore it is necessary to adopt
a suitable model to be able to obtain quantitative results.
Nevertheless, some features of the averaging process, such as the
rapidity dependence in the low-$x$ region, can be appropriately
accounted for by perturbative considerations under the Color Glass
Condensate (CGC) framework \cite{arXiv:1002.0333}.

Regardless of the model under consideration for the specific
calculation of such field averages, the complexity of the
calculation of higher point correlators increases accordingly with
the number of Wilson lines involved. In the same way, the
evolution equations derived perturbatively become very cumbersome
and not well suited for numerical evaluations. The first step on
an attempt to simplify the treatment of such higher point
correlators is to consider the large-$N_c$ limit, in which
the average of a product of traces of Wilson lines reduces to a
product of averages involving only one trace at the time. In
other words, correlations involving fields coming from Wilson
lines from different traces are suppressed by factors of $N_c$.

Now, in this large-$N_c$ limit it becomes necessary to count
appropriately the factors of $N_c$ coming from these multiple
scattering terms and keep only the leading terms in such
expansion. Once all the color matrices (fundamental or adjoint)
from vertices contributions have been removed by means of the
proper color identities, the power of $N_c$ associated to a given
term is just the number of color traces involved.

In principle, all sorts of correlations involving multiple Wilson
lines at various transverse coordinates can appear when
considering the cross section of a multi-particle production
process, but when the large-$N_c$ limit is taken only a few of
those contributions are important. Taking into account the way
that powers of $N_c$ appear from the correlators, it is not
difficult to realize that the terms with simpler structure are the
ones that contribute the most. For a given process, the number and
identity (quark or gluon) of the particles in the final state determines the
maximum number of Wilson lines present in the multiple scattering
terms. The maximum power of $N_c$ present in such terms will
correspond to the configuration in which all the Wilson lines can
be grouped in as many traces as possible, therefore favoring terms
with traces of only a few Wilson lines.

As a first guess, one could suggest that this $N_c$ power counting
implies that the leading $N_c$ contribution always comes from a
term which only includes color dipole amplitudes (traces of two
Wilson lines) and therefore has a maximal number of color traces.
This guess has been proven wrong since thorough studies of
two-particle production processes \cite{JalilianMarian:2004da,Marquet:2007vb,Dominguez:2011wm} have
shown that some processes do not admit a description in terms of
only color dipoles, but that in addition color quadrupoles (traces of four Wilson lines)
are involved as well in the large-$N_c$ limit. This quadrupole amplitude can not in general be written in terms of dipole amplitudes only and its small-$x$ evolution is determined by an independent equation as shown in \cite{JalilianMarian:2004da,Dominguez:2011gc, Dumitru:2010ak}. It has also been shown that there is a direct relation between this quadrupole amplitude and the so-called Weisz\"acker-Williams unintegrated gluon distribution function \cite{Dominguez:2011wm}.

More complicated correlators naturally arise in the calculation of
cross sections of processes with more particles in the final
state, but they are suppressed by powers of $N_c$ as compared with
terms with only dipole and quadrupole amplitudes. They are nevertheless independent from the dipole and quadrupole amplitudes and in principle should be evaluated on their own. A procedure to evaluate higher point correlators, in a Gaussian model in the large-$N_c$ limit, is described in the Appendix and the explicit example of a correlator of six Wilson lines is evaluated explicitly. The small-$x$ evolution of such correlators was recently studied analytically in \cite{Iancu:2011ns} and numerically in \cite{Dumitru:2011vk}, the evaluation shown in the Appendix provides a suitable initial condition for such equations.

The main purpose
of this paper is to show that, in the large-$N_c$ limit, all
multi-particle production processes considered under this
framework can be described in terms of only dipoles and
quadrupoles. We first work out explicitly examples with three particles in the
final state before proceeding to prove the general statement by
induction in the number of particles in the final state. The
inductive step is greatly simplified by the observation that it is
not necessary to consider all the diagrams contributing to a given
process but to find at least one diagram with a term given only by
dipoles and quadrupoles which dominates in the large-$N_c$ limit.

\section{General considerations}\label{gencons}

\begin{figure}
\centering
\subfloat[Sample diagram]{\label{splitplusscat}\includegraphics[width=6cm]{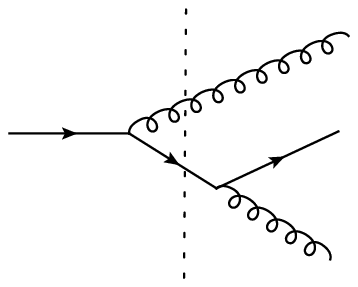}}
\hspace{1cm}
\subfloat[Sample diagram in the large-$N_c$ limit]{\label{splitlargen}\includegraphics[width=6cm]{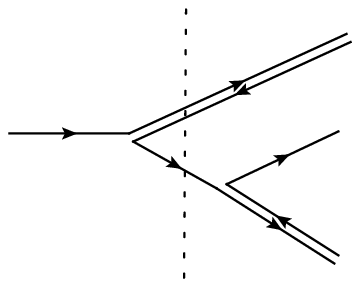}}
\caption{Diagram contributing to the process $q\to qgg$. The dotted line represents the multiple scattering with the target.}
\label{splitdiag}
\end{figure}

For definiteness sake, let us be more specific about the scenario
which we are explicitly considering. We are interested in
processes where multi-particle production takes place in the
presence of a (strong) background field where a highly energetic
initial parton (photon, quark or gluon) undergoes multiple
scatterings. This scenario is particularly well suited for nuclear deep
inelastic scattering (DIS) experiments and forward particle production
in proton-nucleus collisions (using collinear factorization for the proton)
where the high density effects of the target are encoded in the background
field. For this particular case the situation simplifies even more since the
coherence times of the produced particles are long compared with
the length of the target nucleus and therefore the multiple
interaction with the dense system can be considered as
instantaneous. The process is then regarded as an incoming high
energy parton which splits several times into a given final
multi-particle state, interacting at given time with the target
which induces a color rotation in the whole multi-particle system
(see Fig. \ref{splitplusscat}). Of course, one has to consider the
scattering with the target happening at all possible stages of the
splitting process and sum all these contributions, but for the
counting of powers of $N_c$, which is of our interest, this does not
make a difference.

The $N_c$ power counting can only be done at the level of the
cross section after one has already averaged (summed) over initial
(final) colors. Nevertheless, one can make some general
observations at the level of the amplitude based on color
conservation. For a fixed process, with a specific parton in the
initial state and a definite particle content in the final state,
it is possible to determine what is the maximum number of color
traces one can have in the description of the process in the
leading order. Take a diagram for such a process at tree level in
the large-$N_c$ limit replacing all gluon lines with double
quark-antiquark lines which make the color flow explicit, as in
Fig. \ref{splitlargen}. As a consequence of color conservation
each of the external fermion lines must be connected to another
and, since we are considering only tree level diagrams for the
moment, there are no closed loops. From this observation we can see immediately that the maximal number of color traces at tree level is given by the total number of external fermion lines divided by two (with each external gluon contributing two fermion lines). Furthermore, one can see that this maximal number of traces is realized when one considers the square of a given diagram, while interference terms involving different diagrams on the amplitude and conjugate amplitude can possibly have less color traces.

It is also illustrated in Fig. \ref{splitlargen} that each fermion line appears at most twice at the moment of the multiple scattering, regardless of the exact position of the scattering with respect to the splittings. As a consequence, at most two fundamental Wilson lines are color connected at the amplitude level and therefore, when considering the amplitude squared, the respective color traces in the cross section would have only two Wilson lines (dipoles) or four Wilson lines (quadrupoles). The same will be true for any term with the maximal number of traces.

\begin{figure}
\centering
\subfloat[$q\to qg$]{\label{amcon}\includegraphics[width=7cm]{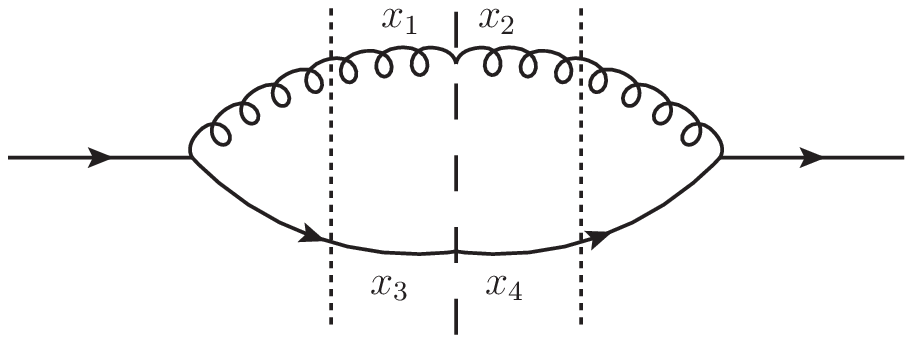}}
\hspace{1cm}
\subfloat[$q\to qg$ in the large-$N_c$ limit]{\label{amconln}\includegraphics[width=7cm]{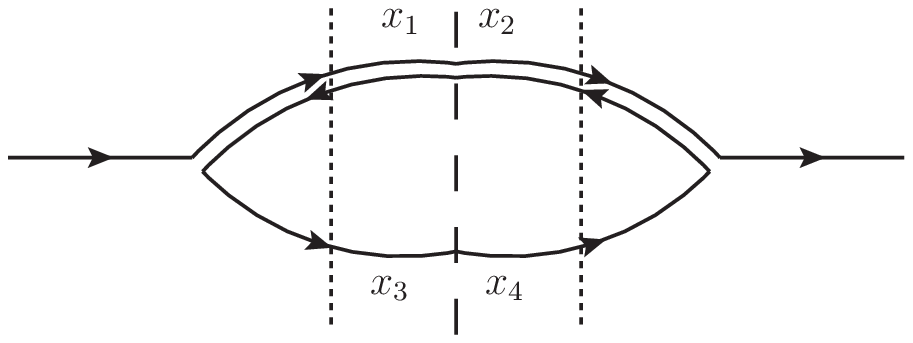}}
\caption{Diagrams with the amplitude and conjugate amplitude for the process $q\to qg$}
\label{amplconjugate}
\end{figure}

Diagrams like those in Fig. \ref{splitdiag} are useful for visualizing the branching of the initial parton into the multi-particle final state but do not contain all the information needed to resolve the color structure of the different contributions to the cross section. For that purpose it is necessary to consider diagrams including the amplitude and the conjugate amplitude where one can perform the necessary color sums and averages. As an example, let us consider the simple process of a quark splitting into a quark and a gluon which then scatters with a background field. Putting the amplitude and the conjugate amplitude in the same diagram we obtain Fig. \ref{amcon} where the dotted lines represent the moment of the scattering in both the amplitude and conjugate amplitude and the dashed line in the middle represents the final state at $t=\infty$. Since the approach employed here makes explicit use of the eikonal approximation to account for the multiple scattering with the external field, it is necessary to consider these diagrams in a coordinate representation where each particle has a definite transverse coordinate. All particles in the final state which are to be detected, and therefore would have a fixed transverse momentum, have different transverse coordinates on each side of the cut at $t=\infty$. When considered in the large-$N_c$ limit, one can see in Fig. \ref{amconln} how some of the fermion lines close into loops, which will contribute one color trace to the cross section, while other lines remain open due to the fact that they contain particles in the initial state. These particles in the initial state should be color connected on both sides of the diagram since one averages over initial colors at the level of the cross section but such connection is not explicit in this kind of diagram.

\begin{figure}[tbp]
\begin{center}
\includegraphics[width=16cm]{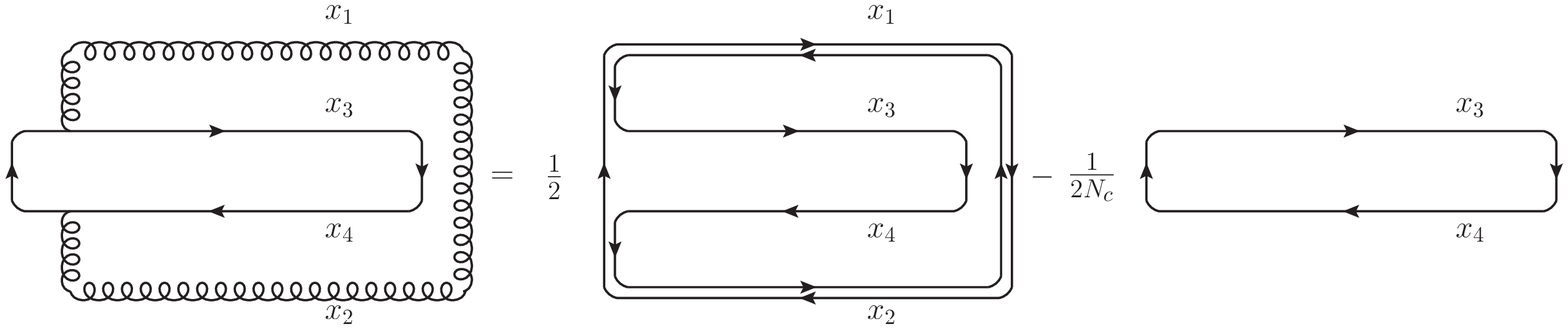}
\end{center}
\caption[*]{Alternative representation for the $q\to qg$.}
\label{qqgg}
\end{figure}

\begin{figure}[tbp]
\begin{center}
\includegraphics[width=8.4cm]{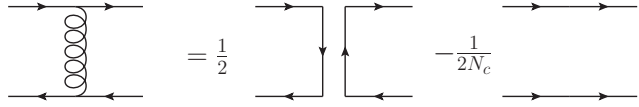}
\end{center}
\caption[*]{Graphical representation of the Fierz identity.}
\label{fi}
\end{figure}

If one is interested in the color structure only, then it is more convenient to draw one fermion loop for each color trace and gluon lines for each adjoint index contraction (including adjoint Wilson lines too). This can be achieved easily by folding on itself the corresponding diagram with the amplitude and conjugate amplitude in such a way that the color connections of the initial state are made explicit. Also, in order to be able to visualize in the diagram which are the Wilson lines entering the expression of the cross section, we will stretch the lines involved in the scattering (both in the amplitude and conjugate amplitude) in the horizontal direction and use the other lines only to illustrate the color connections. For example, the diagram corresponding to Fig. \ref{amcon} is shown in the left hand side of Fig. \ref{qqgg}. One can go one step further and replace all the gluon lines by double fermion lines using the corresponding color identity, which has the graphical representation depicted in Fig. \ref{fi}, in which case the example above takes the form shown in the right hand side of Fig. \ref{qqgg}. Let us focus on the first term, which is the one surviving in the large-$N_c$ limit, where one can identify right away two pieces which are color disconnected, each contributing a color trace to the cross section. One involves two Wilson lines, and therefore is a dipole, while the other involves four, and therefore is a quadrupole.

\begin{figure}
\centering
\subfloat[Dipole amplitude]{\begin{minipage}[c][0.5\width]{0.4\textwidth}
\centering
\includegraphics[width=5cm]{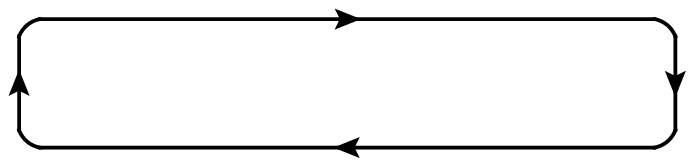}
\end{minipage}}
\hspace{1cm}
\subfloat[Quadrupole amplitude]{\begin{minipage}[c][0.5\width]{0.4\textwidth}
\centering
\includegraphics[width=5cm]{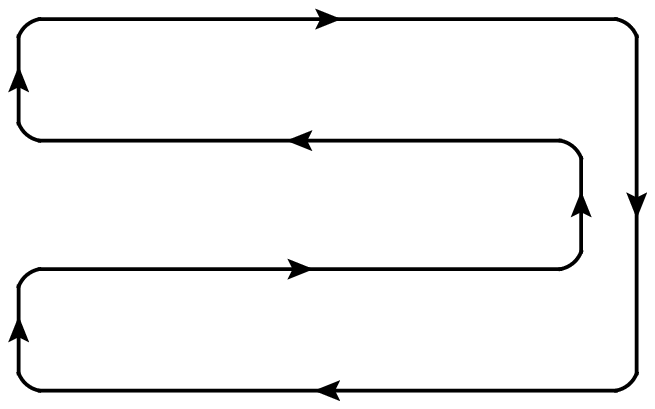}
\end{minipage}}
\caption{Graphical representation of the two amplitudes to be used along the paper.}
\label{dq}
\end{figure}

The example above shows how this graphic approach allows us to recognize easily which kinds of correlators come in the expression for the cross section of a given process. One can also recognize that the combination of Wilson lines describing such multiple scattering is given by
\begin{equation}
\text{Tr}\left[U_3U^\dagger_4t^at^b\right]W^{ac}_1W^{\dagger cb}_2=\frac{1}{2}\text{Tr}\left[U_1U^\dagger_2U_3U^\dagger_4\right]\text{Tr}\left[U^\dagger_1U_2\right]-\frac{1}{2N_c}\text{Tr}\left[U_3U^\dagger_4\right],
\end{equation}
where Fierz identities were used to reach the right hand side. For more complicated processes the color algebra can be very cumbersome and we will rely heavily in the graphic approach to be able to identify the leading $N_c$ piece of the diagrams involved. In particular it will be crucial to identify the pieces corresponding to dipoles and quadrupoles which we show separately on Fig. \ref{dq}.

\begin{figure}
\centering
\subfloat[Interference term]{\label{acint}\includegraphics[width=5cm]{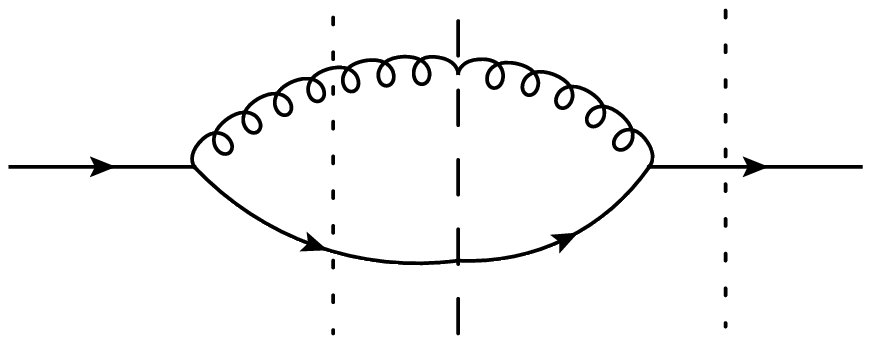}}
\hspace{1cm}
\subfloat[Color structure]{\label{qgqint}\includegraphics[width=9cm]{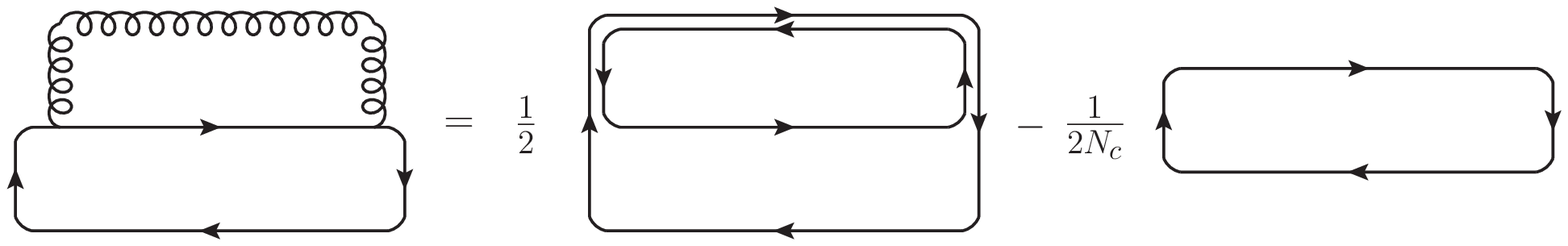}}
\caption{Diagrams for the interference term with interaction after the splitting in the amplitude and before the splitting in the conjugate amplitude.}
\label{qqgint}
\end{figure}

It was also mentioned earlier that one should sum over all the possibilities for the multiple interaction to take place in the amplitude and the conjugate amplitude. Consider the process in the example above but with the multiple interaction occurring in the conjugate amplitude before the splitting. Fig. \ref{qqgint} illustrates how the different diagrams explained above look for this variation of the same process. For this particular case it is clear that the contribution to the cross section can be written in terms of two dipoles. This example illustrates a very important fact that will allow us to concentrate on a smaller number of diagrams: changing the position of the multiple scattering might change the number of Wilson lines appearing in the corresponding term in the cross section but it does not change the number of color traces. It might happen that for some particular cases a color trace is left with no Wilson lines inside, giving a trivial factor of $N_c$, but what is important is that the counting of powers of $N_c$ is the same for a given process regardless of the position of the multiple scattering with respect to the splittings.

The last consideration to be done before starting with the detailed explanation of particular cases is the effect of integrating out particles in the final state. In order to consider inclusive processes it is sometimes necessary to integrate over the momenta of outgoing particles that are not explicitly measured. This happens at the leading order for processes in which one produces an additional fermion in the final state without its corresponding antiparticle, as for example in single inclusive deep inelastic scattering where the incoming photon splits into a quark-antiquark pair but only one of them is measured in the final state. For all other processes this effect comes in at next to leading order and will be fundamental if we want our proof to be complete to all orders.

Under the eikonal approximation at work in this formalism, the multiple scattering terms take the form of simple Wilson lines in the appropriate representation only when considering the process in transverse coordinate space. The momenta of the particles of the final state enters then through a Fourier transform of the respective coordinate in the amplitude and conjugate amplitude. Therefore integrating over the transverse momentum of a particle in the final state gives a delta function which identifies the transverse coordinate of the particle in the amplitude with the one in the conjugate amplitude. Given the unitarity of the Wilson lines, this sort of manipulation will likely reduce the number of Wilson lines appearing in the corresponding contribution to the cross section of any particular diagram. The sum over color indices in the final state guaranties that if a final state particle participates in the multiple scattering both in the amplitude and the conjugate amplitude, both Wilson lines appear next to each other in the contribution to the cross section, and if they are placed at the same transverse coordinate then they exactly cancel out. This sort of cancellation between real and virtual contributions for final state interactions has been studied before and it is well known to be an immediate consequence of unitarity \cite{Mueller:2001fv}, the only place where interactions with the unmeasured particle come in is in the interference terms between initial and final state interactions.

From the point of view of the correlators we are interested in, it is clear that the ones entering the cross sections of these more inclusive processes are either the same or simpler than the ones involved in the expression for the full exclusive process where all particles in the final state are detected. This observation allows us to focus our attention to the fully exclusive processes only.

\section{Known cases: one and two particles in the final state}

Since our goal is to prove by induction an statement that will be valid for all multi-particle production processes in the previously described setup, it is necessary to start our analysis with the simplest cases available. All of these cases have been previously studied in the literature, here we comment in the results and emphasize the features of the derivations which will be useful for the development of the argument for general cases. 

We start with processes with one particle in the final state even though their color structure is better understood in the context of processes with two particles in the final state. The reason for this is that the non-trivial contributions to these processes are obtained after integrating out one of the particles in the final state. As explained in the previous section, this yields procedure gives way to simpler expressions where the correlators appearing in the cross section have less Wilson lines than the corresponding correlators for the processes with two final state particles. The choice to present first the one-particle processes follows the thread of the main idea where we intend to organize the processes in terms of the number of particles in the final state, from there on the emphasis will be put on fully exclusive processes with less exclusive processes already accounted for by the observations of the previous section.

\subsection{SIDIS}\label{sidis}

Semi-inclusive deep inelastic scattering (SIDIS) has been widely studied in the literature since it has been recognized to give access to transverse momentum dependent parton distribution functions. In the context of saturation physics, several studies have cemented the foundations of the formalism to treat deep inelastic scattering processes where the focus was mainly on the total cross section and form factors. Most of these studies are based on the dipole model approach \cite{CU-TP-441a}, which shows directly a clear relation between the total cross section and the forward amplitude of a color dipole going through a background color field, but the same conclusions can be found from the setup explained in the previous sections after one integrates out all the particles in the final state. For the particular case of SIDIS, it was explicitly shown in \cite{McLerran:1998nk,Kovchegov:1999kx,Mueller:1999wm} that the only correlator needed in the expression of the cross section is the dipole amplitude, and in \cite{Marquet:2009ca} a direct connection to the transverse momentum dependent quark distributions was made.

The lowest order calculation of this process is very simple from the point of view of the multiple scattering, as we are interested in here. It can be shown that for a DIS process, in which one considers a virtual photon splitting into a quark-antiquark pair, the multiple scattering term at the amplitude level always appears as $1-U(x_1)U^\dagger(x_2)$, where $x_1$ and $x_2$ are the transverse positions of the quark and antiquark respectively. Clearly, the square of this amplitude will have terms with traces of either zero, two, or four Wilson lines, but since one integrates over the momentum of one of the particles, two of the Wilson lines are at the same transverse coordinate and therefore cancel out in the term with four Wilson lines.

We note that, by swapping the initial state photon and the quark (or antiquark) in the final state whose momentum is integrated out, the SIDIS process turns into photon+hadron production in p+A collisions (whether the photon is measured or not doesn't matter because it does not multiply scatter with the gluons of the target). Therefore the only correlator needed to express of that cross section is also the dipole amplitude.

\subsection{Single hadron production in $pA$}

This process is of particular importance in the context of the study of cold nuclear matter effects. One of the first measurements to show deviations from the purely additive scheme where a nucleus is considered as an uncorrelated ensemble of nucleons was the $p_\perp$-spectrum of produced hadrons in a proton(deuteron)-nucleus collision. Several measurements of the nuclear modification factor for inclusive hadron production were performed showing a clear enhancement at mid-rapidities and suppression at forward rapidities  \cite{Arsene:2004ux,Adams:2006uz}, which cannot be understood without coherent scattering involving several participants in the nucleus.

In the framework described in this paper, the lowest order contribution to this process is straightforward to calculate as the convolution of a quark distribution for the projectile with a dipole amplitude formed by the Wilson lines corresponding to the quark in the amplitude and the conjugate amplitude. At the partonic level this is simply transverse momentum broadening of a quark going through a nucleus \cite{Dumitru:2002qt}. This level of the calculation has been proven to not be enough to describe the available data and therefore it is necessary to consider additional contributions to the cross section coming mainly from gluon emissions, with the additional complication of having more particles in the final state. As a first attempt, it was shown in \cite{Kovchegov:1998bi,Dumitru:2001ux} that one can easily calculate the soft gluon limit where the longitudinal momentum of the gluon is much smaller than that of the original quark and, therefore, the parent quark does not feel any recoil effect after the emission. This no-recoil condition is present in the transverse coordinate space formalism in the form of the parent quark having the same transverse coordinate before and after the emission of the gluon, allowing an easy way to rewrite the interference terms between scattering before and after the splitting in a convenient way in terms of only gluon dipoles \cite{Kovner:2001vi,Kovchegov:2001sc,Marquet:2004xa}. The details of how this manipulation works out in the Wilson line language will be postponed to the section on di-jet production.

The soft gluon approximation implies a large rapidity gap between the measured hadron and the remnants of the proton in the forward region. In order to extend the region where the calculation is applicable, and in particular include hadrons in the forward region where the effects of saturation are expected to be stronger, it is necessary to consider the full vertex and allow the parent quark to have different transverse positions before and after the emission. This was first done in \cite{Dumitru:2005gt} where the emission vertex is treated exactly but only the divergent pieces are kept after integrating over the momentum of one of the final state particles. These divergent pieces are shown to be absorbed by the parton distribution functions and fragmentation functions attached to the initial and final state partons by considering the fully DGLAP evolved distributions. In Ref.~\cite{Altinoluk:2011qy}, an additional contribution to the cross-section, formally of next-to-leading order but nevertheless important to restore the correct high$-p_T$ limit, was calculated. The complete NLO calculation, including the calculation of all virtual terms and also the finite terms after integration over one of the outgoing momenta, was recently done in \cite{Chirilli:2011km}. In all cases it can be seen that, in the large-$N_c$ limit, only dipole amplitudes are needed in the full expression for the cross section. In terms of the small-$x$ evolution at NLO, the same conclusion also holds \cite{Mueller:2012bn}.  In additional, it has been demonstrated in Ref.~\cite{Mueller:2012bn} that the dipole formalism employed in this calculation holds at next-to-leading order accuracy. 

As a matter of fact, one can use induction to prove that, for any single inclusive processes in terms of any order in $\alpha_s$, the scattering amplitudes only contain dipole amplitudes in the large $N_c$ limit. As discussed earlier, the above conclusion holds up to LO and NLO for $pA$ collisions. To obtain the contribution at NNLO, one just needs to add one more gluon with one single coordinate among the dipoles at the previous order (NLO). Using the Fierz identity, one can easily prove that all the relevant graphs can be reduced to products of dipole amplitudes at NNLO in the large $N_c$ limit. The proof for single inclusive DIS productions is identical. Therefore, we can conclude that all single inclusive processes can be universally described by the dipole amplitudes at the large $N_c$ limit. From the universality point of view, the large $N_c$ limit plays an important and indispensable role for the factorization proof. As shown in Refs~\cite{Chirilli:2011km}, higher point functions, which are new objects, contribute to the cross sections as large $N_c$ corrections. Without the large $N_c$ limit, there is no universality, hence no factorization for single inclusive processes. 

\subsection{Di-jet production in DIS}\label{secdijetdis}

The process of di-jet production in DIS is of particular interest to us since it is the simplest process where the quadrupole is needed for an accurate description of the multiple scattering factors entering the expression for the cross section. It was carefully studied in \cite{Dominguez:2010xd, Dominguez:2011wm} where it was also emphasized its direct relation to the Weizs\"acker-Williams gluon distribution function.

\begin{figure}
\centering
\subfloat[Amplitude squared]{\label{djdissq}\includegraphics[width=7cm]{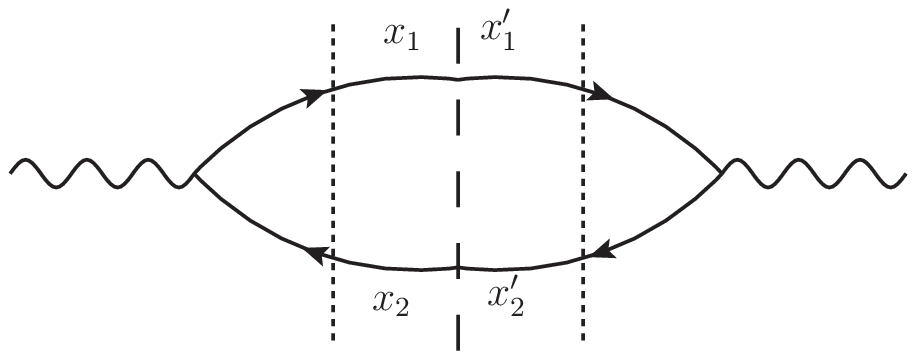}}
\hspace{1cm}
\subfloat[Color structure]{\label{qdis}\includegraphics[width=5cm]{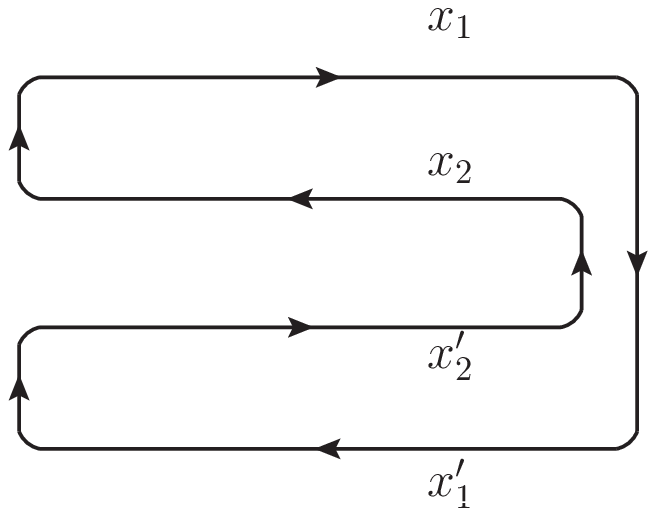}}
\caption{Diagrams for the quadrupole term in the di-jet production in DIS. The photon is omitted in (b) since we are interested only in the color structure.}
\label{dijetdisfigs}
\end{figure}

It was already mentioned in Section \ref{sidis} and previously in \cite{Dominguez:2010xd, Dominguez:2011wm, Gelis:2002nn} that, to leading order at the amplitude level, the corresponding multiple scattering factor for a process including a photon splitting into a quark-antiquark pair is given by $1-U(x_1)U^\dagger(x_2)$, where $x_1$ and $x_2$ are the transverse positions of the quark and antiquark respectively. Since for the di-jet process one is interested in keeping explicit the momentum variables for both particles, the transverse coordinates in the amplitude and conjugate amplitude are different and therefore there is a term with four Wilson lines in the expression for the cross section. This quadrupole term takes the form $\left\langle\text{Tr}\left[U(x_1)U^\dagger(x_2)U(x'_2)U^\dagger(x'_1)\right]\right\rangle$ and clearly corresponds to the contribution from the diagram where the interaction with the background field occurs after the splitting both in the amplitude and conjugate amplitude. Fig. \ref{dijetdisfigs} shows the two ways of graphically representing this process as indicated in Section \ref{gencons}.

\subsection{Di-jet production in $pA$}

Processes with two particles in the final state of a proton-nucleus collision have become increasingly important in the last few years given the availability of new data and the unique character of the physics that can be probed through these particular sort of measurement. By constraining the kinematics of the two outgoing particles, one can, at leading order, separately fix the longitudinal momentum fractions carried by the incoming partons (up to the additional fragmentation integrals involved in the case of di-hadron production), something that cannot be achieved by one-particle measurements. This advantage makes this kind of process very attractive to try to measure high-density effects characteristic of the small-$x$ part of the target wave function, and in fact the RHIC measurement of di-hadron correlations in the forward region \cite{Braidot:2010zh,Adare:2011sc} is considered as the strongest evidence to date of saturation.

Several studies concerning this kind of process are available in the literature, those include the role of quark distributions \cite{Xiao:2010sp}, general descriptions on how to calculate the cross sections in the presence of background fields \cite{JalilianMarian:2004da,Nikolaev:2005dd,Baier:2005dv,Marquet:2007vb}, more phenomenological applications including small-$x$ evolution which make direct contact with data \cite{Albacete:2010pg,Stasto:2011ru,Lappi:2012nh}, and factorization studies where a direct relationship is established between these observables and unintegrated gluon distributions \cite{Dominguez:2011wm}. Here we will just present a summary of the correlators appearing in the cross section for each of these processes and the specific form they take in the large-$N_c$ limit.

The first case to be considered is the quark initiated process where a quark from the projectile splits into a quark and a gluon which are both detected in the final state. This is precisely the process chosen in Section \ref{gencons} as illustration of the sort of analysis to be performed throughout this paper. There it was shown how all the terms entering the cross section for that process at leading order in the large-$N_c$ limit involve only the dipole and quadrupole correlators.

Now we turn our attention to processes with gluons in the initial state. The first one to be considered is the case where an initial gluon from the projectile splits into a quark-antiquark pair. The color algebra for this process in the large-$N_c$ limit is particularly simple since the cross section can be written in terms of only dipole correlators, nevertheless, it is important to look closely at some of the aspects of this process since it will allow us to draw general conclusions about any process with quark-antiquark pairs in the final state. The key aspect here is that the splitting does not introduce any additional color flow in the large-$N_c$ limit, in the double line notation introduced for the large-$N_c$ limit the only effect of such vertex is to separate the two lines and allow for different transverse coordinates for the quark and the antiquark. Such separation has no consequences in the trace structure of the leading part of the diagram when written entirely in terms of fundamental Wilson lines.

\begin{figure}[tbp]
\begin{center}
\includegraphics[width=16cm]{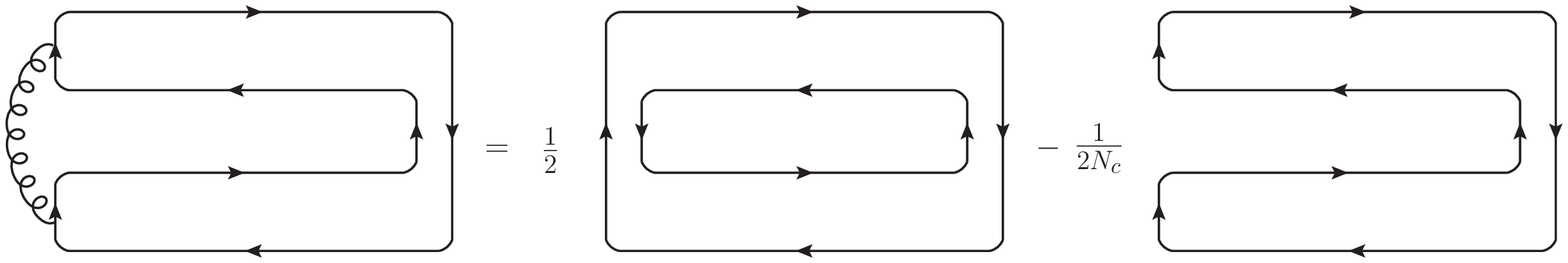}
\end{center}
\caption[*]{Color structure of $g\to q\bar{q}$ process with interaction after the splitting.}
\label{gqqafter}
\end{figure}

The observation above can be easily illustrated when one compares the different contributions to the $g\to q\bar{q}$ process arising from having the multiple scattering before or after the splitting. For definiteness, consider first the case where the scattering occurs after the splitting both in the amplitude and in the conjugate amplitude. The partons involved in the scattering are therefore two quark-antiquark pairs as indicated on the left hand side of Fig. \ref{gqqafter}. This multiple scattering term is written in terms of Wilson lines (after averaging over the initial color of the gluon) as $\text{Tr}\left[U_1t^aU^\dagger_2U_3t^aU^\dagger_4\right]$. Using the Fierz identity to get rid of the color matrices one obtains
\begin{equation}
\text{Tr}\left[U_1t^aU^\dagger_2U_3t^aU^\dagger_4\right]=\frac{1}{2}\text{Tr}\left[U_1U^\dagger_4\right]\text{Tr}\left[U^\dagger_2U_3\right]-\frac{1}{2N_c}\text{Tr}\left[U_1U^\dagger_2U_3U^\dagger_4\right].
\end{equation}
This relation is represented graphically in Fig. \ref{gqqafter}. Now consider the case where the interaction occurs after the splitting in the amplitude and before the splitting in the conjugate amplitude. Then, the partons involved in the scattering are a quark-antiquark pair and a gluon and the multiple scattering factor, in terms of the corresponding Wilson lines, is
\begin{equation}
\text{Tr}\left[U_1t^aU^\dagger_2t^b\right]W^{ba}_3=\frac{1}{2}\text{Tr}\left[U_1U^\dagger_3\right]\text{Tr}\left[U^\dagger_2U_3\right]-\frac{1}{2N_c}\text{Tr}\left[U_1U^\dagger_2\right].
\end{equation}
It is easy to see that the graphical representation of the color structure of this process is topologically equivalent to that shown in Fig. \ref{qgqint}. As previously anticipated, the message to take from here is that the color flow structure present in this process is the same for both cases presented above for the leading part in the large-$N_c$ limit. Moreover, this is the same structure we see when only two gluons are present in the multiple scattering, always the product of two fundamental dipoles. As opposed to processes where gluons are emitted, no extra color charge is created at the vertex and therefore the color structure remains unchanged, the same is also true for processes with a quark-antiquark pair merging into a gluon which might be important for higher order contributions.

The observations above allows us to neglect, from the point of view of the color structure of the process in the large-$N_c$ limit, all the vertices involving a gluon splitting into a quark-antiquark pair and focus our attention in processes where all the vertices involve a quark emitting a gluon or a gluon splitting into two gluons. The four-gluon vertex can also be ignored since its color structure is equivalent to a combination of three-gluon vertices.

Up until now, the large-$N_c$ limit has been mainly invoked to be able to regard color traces as separate entities, even after one takes the average over the background color field, arguing that correlations between fields entering Wilson lines in different color traces are suppressed by factors of $1/N_c^2$. The multiple scattering factors of the processes considered so far, when expressed in terms of fundamental Wilson lines only, can all be expressed in terms of products of traces of either two or four Wilson lines in the fundamental representation. This will not be the case from now on when we start to consider more complicated processes, starting with the case of an initial gluon splitting into two gluons which fragment independently. This process will have contributions from diagrams with three and four gluons present at the moment of the multiple scattering which will lead to traces of six and eight fundamental Wilson lines.

\begin{figure}[tbp]
\begin{center}
\includegraphics[width=0.8\textwidth]{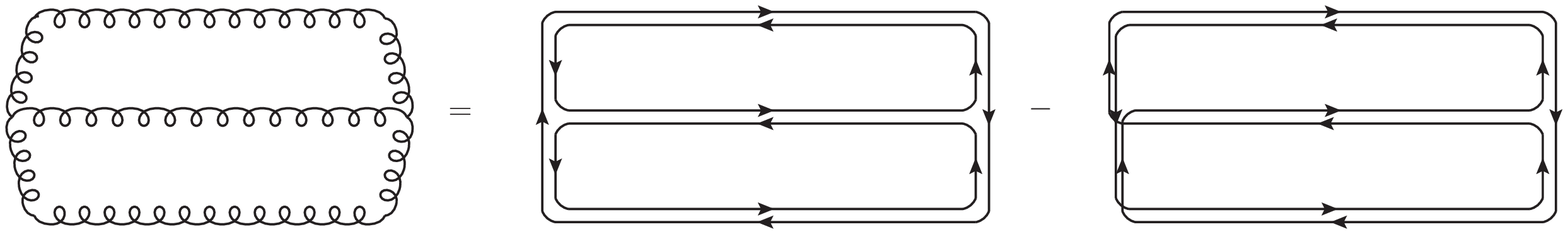}
\end{center}
\caption[*]{The graphs on the right side the equation should also include
their Hermitian conjugates which can be obtained by just simply reversing
all the arrows. Here all the correlators are assumed to be real, therefore
the $\frac{1}{2}$ factor is cancelled.}
\label{gggint}
\end{figure}

\begin{figure}[tbp]
\begin{center}
\includegraphics[width=8.4cm]{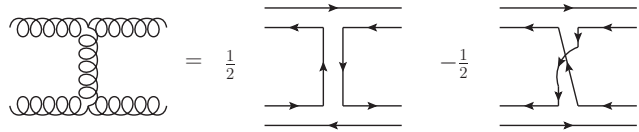}
\end{center}
\caption[*]{Double line representation of the gluon vertex. Here we should
include the Hermitian conjugates of the graphs on the right hand side of the
equation.}
\label{gv}
\end{figure}

Let us consider first the case where the multiple scattering occurs after the splitting in the amplitude and before the splitting in the conjugate amplitude. It is easy to see that there are three gluons involved in the multiple scattering, each one contributing one adjoint Wilson lines to the multiple scattering factor, and that they are connected before and after the scattering by three-gluon vertices. Following Ref.~\cite{Dominguez:2011wm}, we can evaluate the relevant scattering matrices which yield 
\begin{eqnarray}
&&f_{ade}W_{1}^{db}W_{2}^{ec}f_{fbc}W_{3}^{af}  \notag \\
&=&\frac{1}{2}\text{Tr}\left[ U_{1}U_{3}^{\dagger }\right] \text{Tr}\left[
U_{3}U_{2}^{\dagger }\right] \text{Tr}\left[ U_{2}U_{1}^{\dagger }\right] +%
\frac{1}{2}\text{Tr}\left[ U_{2}U_{3}^{\dagger }\right] \text{Tr}\left[
U_{1}U_{2}^{\dagger }\right] \text{Tr}\left[ U_{3}U_{1}^{\dagger }\right]
\notag \\
&&-\frac{1}{2}\text{Tr}\left[ U_{3}U_{2}^{\dagger }U_{1}U_{3}^{\dagger
}U_{2}U_{1}^{\dagger }\right] -\frac{1}{2}\text{Tr}\left[ U_{1}U_{2}^{%
\dagger }U_{3}U_{1}^{\dagger }U_{2}U_{3}^{\dagger }\right] .\label{wlgggint}
\end{eqnarray}
The corresponding graphical representation is shown in Fig. \ref{gggint}. Finding the correct graphical representation, in terms of fermion lines only, without performing explicitly the algebra shown in Eq. (\ref{wlgggint}) is possible as long as one knows how to represent two consecutive three-gluon vertices in the double line representation. The identity to be used is illustrated in Fig. \ref{gv}.

Both Eq. (\ref{wlgggint}) and Fig. \ref{gggint} show that the term with a trace of six Wilson lines (sextupole) is suppressed in the large-$N_c$ limit. On one hand, the terms with a product of three color traces are proportional to $N_c^3$ while the terms with only one trace are of order $N_c$. On the other hand, the graphical representation clearly shows that the term with the sextupole comes from a nonplanar diagram and therefore is suppressed with respect to planar diagrams by at least a factor of $1/N_c^2$.

\begin{figure}[tbp]
\begin{center}
\includegraphics[width=0.8\textwidth]{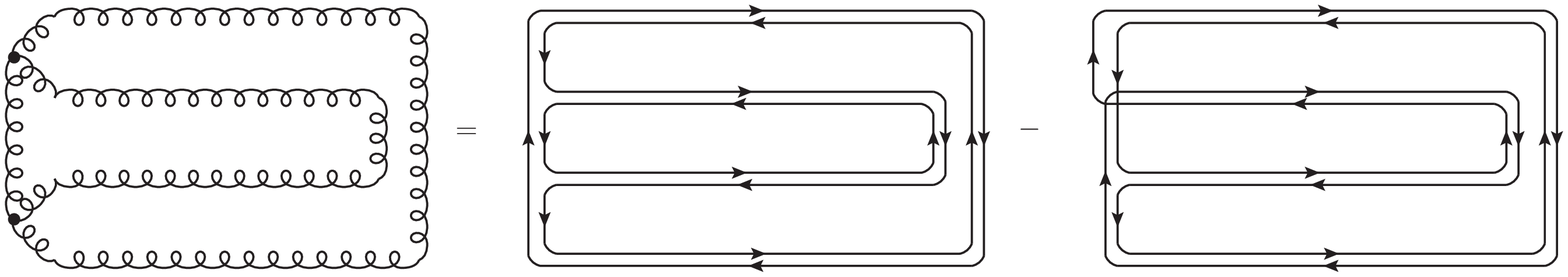}
\end{center}
\caption[*]{The graphs on the right side the equation should also include
their Hermitian conjugates which can be obtained by just simply reversing
all the arrows. Here all the correlators are assumed to be real, therefore
the $\frac{1}{2}$ factor is cancelled.}
\label{gggsq}
\end{figure}

The case where the multiple scattering occurs after the splitting, both in the amplitude and the conjugate amplitude, is treated similarly. Now, there is an extra gluon present in the multiple scattering, contributing an extra adjoint Wilson line, and the two three-gluon vertices are placed before the scattering. The color connections after the scattering are given by identifying the corresponding gluons from the amplitude and the conjugate amplitude. In summary, the relevant multiple scattering term, in terms of adjoint Wilson lines, as well as fundamental Wilson lines, is
\begin{eqnarray}
&&f_{ade}\left( W_{1}W_{2}^{\dagger }\right) ^{db}f_{abc}\left(
W_{3}W_{4}^{\dagger }\right) ^{ec}  \notag \\
&=&\frac{1}{2}\text{Tr}\left[ U_{2}U_{1}^{\dagger }\right] \text{Tr}\left[
U_{3}U_{4}^{\dagger }\right] \text{Tr}\left[ U_{1}U_{2}^{\dagger
}U_{4}U_{3}^{\dagger }\right] +\frac{1}{2}\text{Tr}\left[ U_{4}U_{3}^{%
\dagger }\right] \text{Tr}\left[ U_{1}U_{2}^{\dagger }\right] \text{Tr}\left[
U_{3}U_{4}^{\dagger }U_{2}U_{1}^{\dagger }\right]  \notag \\
&&-\frac{1}{2}\text{Tr}\left[ U_{2}U_{1}^{\dagger }U_{3}U_{4}^{\dagger
}U_{1}U_{2}^{\dagger }U_{4}U_{3}^{\dagger }\right] -\frac{1}{2}\text{Tr}%
\left[ U_{1}U_{2}^{\dagger }U_{3}U_{4}^{\dagger }U_{2}U_{1}^{\dagger
}U_{4}U_{3}^{\dagger }\right] .\label{wlgggsq}
\end{eqnarray}
Its graphical representation is given in Fig. \ref{gggsq}, where once again we made use of the identity illustrated in Fig. \ref{gv}. Similarly to the previous case, it is easy to see that the term with three color traces is the one that dominates in the large-$N_c$ limit, and the term with a trace of eight Wilson lines (octupole) is subleading. We can then safely state that, for all the cases considered in this section, the leading contribution to the multiple scattering term in the large-$N_c$ limit can always be written in terms of dipole and quadrupoles correlators only.

\section{Three particles in the final state}\label{sec:3part}

The aim of this section is to show with explicit examples how the color structure of a given process is affected by the inclusion of additional gluons in the final state. At the end of the previous section it was seen already that when the complexity of the problem increases, as well as the number of colored particles participating in the multiple scattering, it is natural that new higher point correlations have to be included in the full description of the process. It was also seen that these higher point correlations appear always suppressed by inverse powers of $N_c$ when the multiple scattering factor is expressed in terms of only fundamental Wilson lines. The same will be observed in the examples shown in this section, higher point correlations keep appearing but the leading term in the large-$N_c$ limit can always be written in terms of dipole and quadrupole amplitudes.

\begin{figure}[tbp]
\begin{center}
\includegraphics[width=0.8\textwidth]{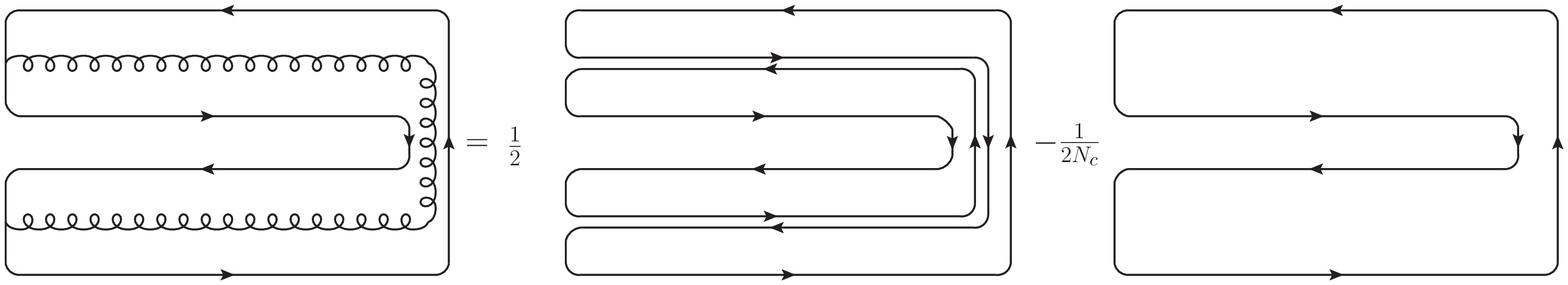}
\end{center}
\caption[*]{$q\bar{q}g$ production in DIS.}
\label{qqg}
\end{figure}

Let us start with one of the simplest cases at this level, which at the same time will show us a general feature of further inclusion of gluons. Consider the process of production of a quark-antiquark pair and a gluon in DIS. As was observed in previous cases, it is sufficient to consider the case where the maximum number of particles participate in the multiple scattering since any other combination would yield  simpler correlators. The process is then described by a photon splitting into a quark-antiquark pair, which emits a gluon before undergoing multiple scatterings with the target field both in the amplitude and conjugate amplitude. The multiple scattering term includes therefore two quark-antiquark pairs and two gluons with the color connections as shown in Fig. \ref{qqg}, and which can be written as

\begin{eqnarray}
&&\text{Tr}\left[ U_{1}^{\dagger }t^{a}U_{2}U_{4}^{\dagger }t^{b}U_{3}\right]
\left( W_{5}W_{6}^{\dagger }\right) ^{ab}  \notag \\
&=&\frac{1}{2}\text{Tr}\left[ U_{2}U_{4}^{\dagger }U_{6}U_{5}^{\dagger }%
\right] \text{Tr}\left[ U_{3}U_{1}^{\dagger }U_{5}U_{6}^{\dagger }\right] -%
\frac{1}{2N_{c}}\text{Tr}\left[ U_{3}U_{1}^{\dagger }U_{2}U_{4}^{\dagger }%
\right] ,
\end{eqnarray}

Both graphically and algebraically it is easy to see that the effect of adding a new gluon to the process studied in Section \ref{secdijetdis} can be understood in terms of the Fierz identity. Here we choose to interpret the result in terms of the graphical representation. The extra gluon can be removed from the diagram by the rule depicted in Fig. \ref{fi}, regardless of it being involved in the multiple interaction. For this particular case, the first term in the right hand side of Fig. \ref{fi} cuts in two the already existing quadrupole, giving as a result a term with two color traces which will clearly dominate in the large-$N_c$ limit as compared to the second term which does not introduce any new traces and has an extra factor of $1/N_c$ in front. Here the original quadrupole is split into two quadrupoles, but it is easy to see that for slightly different cases, as for example if one considers the term with scattering before the emission of the gluon in the conjugate amplitude, one can end up with a quadrupole and a dipole. These small variations of the same process can be seen as attaching the gluon to different places of the original quadrupole. At the end of the day the result is always the same, the leading term in the large-$N_c$ limit will be given by the term that splits the original color trace into two different color traces.

\begin{figure}
\centering
\subfloat[]{\includegraphics[width=5cm]{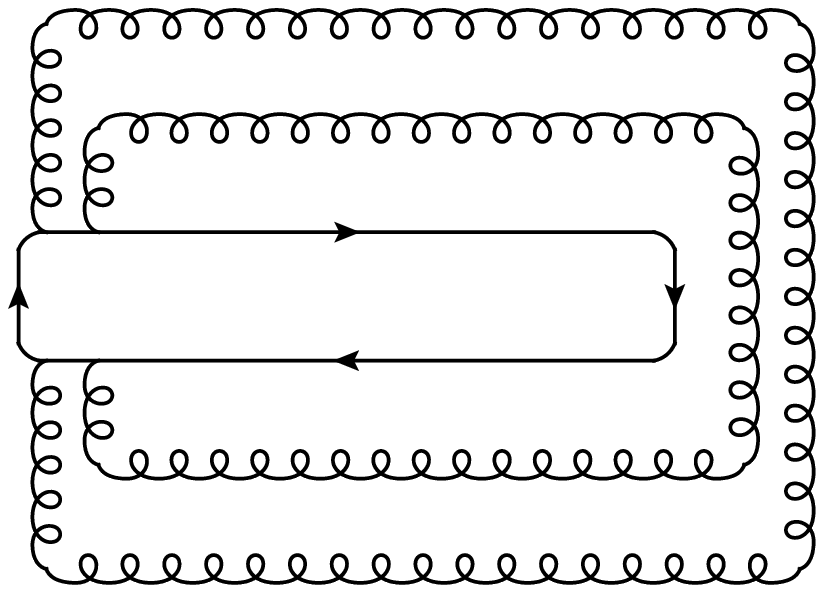}}
\hspace{1cm}
\subfloat[]{\includegraphics[width=5cm]{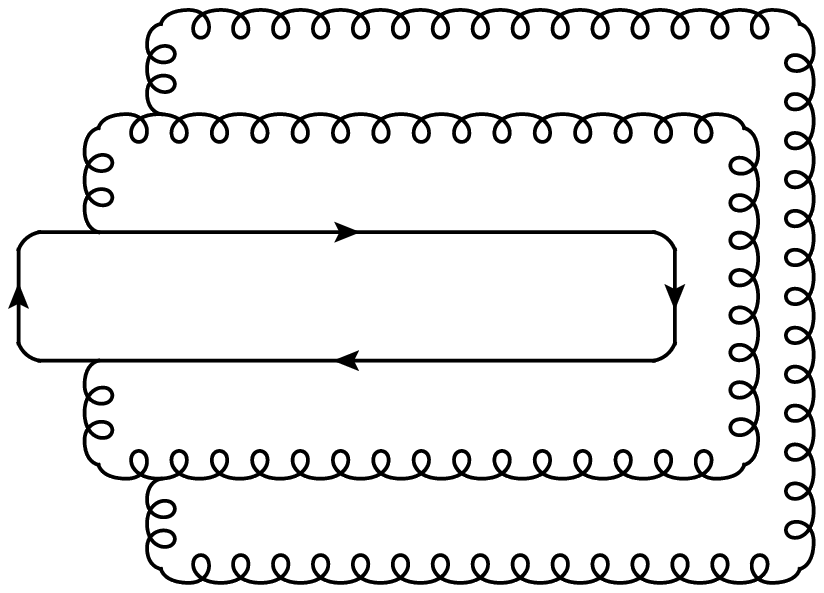}}
\caption{$q\to qgg$}
\label{qgg}
\end{figure}

\begin{figure}[tbp]
\begin{center}
\includegraphics[width=0.8\textwidth]{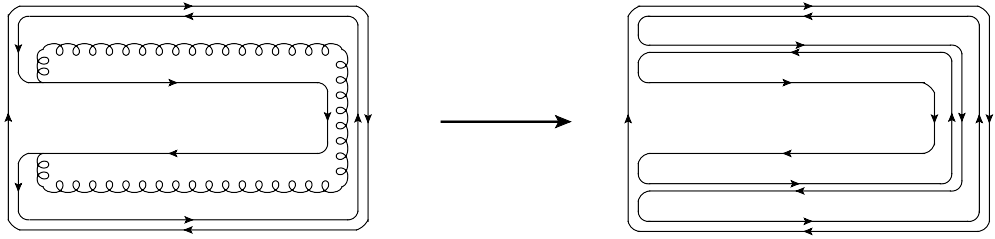}
\end{center}
\caption[*]{Illustration of the effect of the extra gluon in the leading piece for one diagram contributing to the $q\to qgg$ process.}
\label{qdtoqqd}
\end{figure}

Next, we switch our attention to processes present in $pA$ collisions. First consider the quark initiated process with one quark and two gluons in the final state. The relevant diagrams can all be obtained by considering all possible ways of attaching one extra gluon to the diagrams contributing to the $q\to qg$ process. As an illustration, consider the diagrams corresponding to the square of the processes where the additional gluon is emitted either by the final quark or by the first gluon and all three particles participate in the interaction with the background field. The color structure is given by the diagrams of Fig. \ref{qgg}. \footnote{Here we are only interested in the planar diagrams, in which gluon lines do not cross each others. The non-planar diagrams are large $N_c$ suppressed, and thus are always discarded.} It is easy to see that for the case where the second gluon is emitted from the quark the color structure can be resolved by a straightforward application of the Fierz identity, leading to the conclusion that the leading term in the large-$N_c$ limit is given by one dipole and two quadrupoles. Instead of getting into the algebraic details of this computation, we choose the graphical approach here to show how the additional gluon modifies the structure of the correlator. Consider the diagram in the first term of the right hand side of Fig. \ref{qqgg}, representing the leading piece of the $q\to qg$ process, and include the additional gluon as shown in Fig. \ref{qdtoqqd}. Since the gluon is only attached to the quadrupole part one can momentarily forget about the dipole part, it is clear that when one uses the Fierz identity to remove the gluon from the diagram one obtains that the dominant piece of the diagram is the one in which the quadrupole is split into two quadrupoles.

For the case just described, there was no other choice but to attach the additional gluon to the quadrupole from the leading piece of the $q\to qg$ process. When one considers other diagrams, as the one already described where the second gluon is emitted from the first gluon, there are other ways of attaching this new gluon to the dominant part of the diagram. In general one can consider all possible ways of attaching the two legs of the new gluon and quickly realize that the pieces which will survive in the large-$N_c$ limit are the ones coming from attaching both legs of the new gluon to the same fermion loop. As an illustration, let us consider the algebraic expression of the scattering term for the diagram in Fig. \ref{qgg}(b),
\begin{eqnarray}
&&\text{Tr}\left[ U_{5}^{\dagger }t^{a}t^{a^{\prime }}U_{6}\right]
f_{ade}\left( W_{1}W_{2}^{\dagger }\right) ^{db}f_{a^{\prime }bc}\left(
W_{3}W_{4}^{\dagger }\right) ^{ec}  \notag \\
&=&\frac{1}{4}\text{Tr}\left[ U_{1}U_{2}^{\dagger }U_{4}U_{3}^{\dagger }
\right] \text{Tr}\left[ U_{6}U_{5}^{\dagger }U_{3}U_{4}^{\dagger }\right]
\text{Tr}\left[ U_{2}U_{1}^{\dagger }\right]   \notag \\
&&+\frac{1}{4}\text{Tr}\left[ U_{3}U_{4}^{\dagger }U_{2}U_{1}^{\dagger }
\right] \text{Tr}\left[ U_{6}U_{5}^{\dagger }U_{1}U_{2}^{\dagger }\right]
\text{Tr}\left[ U_{4}U_{3}^{\dagger }\right]   \notag \\
&&-\frac{1}{4}\text{Tr}\left[ U_{6}U_{5}^{\dagger }U_{1}U_{2}^{\dagger
}U_{4}U_{3}^{\dagger }U_{2}U_{1}^{\dagger }U_{3}U_{4}^{\dagger }\right]
\notag \\
&&-\frac{1}{4}\text{Tr}\left[ U_{6}U_{5}^{\dagger }U_{3}U_{4}^{\dagger
}U_{2}U_{1}^{\dagger }U_{4}U_{3}^{\dagger }U_{1}U_{2}^{\dagger }\right] .
\end{eqnarray}
One can clearly see that the two first terms come from attaching the two legs of the new gluon to the same preexisting fermion loop while the last two terms come from attaching the two legs of the new gluon to different fermion loops. Higher point correlators appear in this expression but always suppressed by a power of $1/N_c^2$ as compared to the terms with only dipoles and quadrupoles.

For the gluon initiated processes the situation is very similar. Even though the explicit calculations are more intricate and tedious, one can easily recognize that the leading terms for large-$N_c$ for the process $g\to ggg$ consist of a combination of two dipoles and two quadrupoles. One can perform a similar analysis to the one performed above for the $q\to qgg$ process starting with the leading piece of the $g\to gg$ process and adding an extra gluon. Again, the terms that survive in the large-$N_c$ limit will be the ones coming from attaching both legs of the new gluon to only one of the preexisting fermion loops, in this case two dipoles and one quadrupole, creating therefore one extra quadrupole (in the case where all produce particles participate in the multiple scattering both in the amplitude and conjugate amplitude).

\section{Proof of the general case}

The argument used in the previous section to go from processes with two particles in the final state to processes with three particles in the final state can be generalized in a straightforward manner to an arbitrary number of particles in the final state. In order to do so in a consistent matter we will show by induction that adding a new particle to the final state does not change the fact that the leading terms in the large-$N_c$ limit are given in terms of only dipoles and quadrupoles. We find it convenient to focus first on the leading order for a given number of particles, then we show how the argument can be further generalized to include also all order contributions.

\subsection{Leading order}

Before getting into the details of the inductive step of the proof, let us summarize some of the observations that have been done through the paper which will be useful for the argument presented in this section.

\begin{enumerate}
\item In the large-$N_c$ limit, the $N_c$ power counting is most easily done when scattering factors are expressed fully in terms of fundamental Wilson lines, where each trace contributes with a factor of $N_c$.
\item Changing the moment of the multiple interaction does not change the $N_c$ power counting. The case where all the final state particles participate in the multiple interaction is where the most complicated correlators can possibly be found and therefore these will be the only cases under study in this section.
\item  A gluon splitting into a quark-antiquark pair does not add any color charge to the process and therefore leaves its color structure unchanged. For the argument presented here one can neglect any new contributions from processes with extra quark-antiquark pairs in the final state and always assume that the additional particles are gluons.
\item From the point of view of the color structure only, a four-gluon vertex can always be represented by the sum of different ways of combining two three-gluon vertices. Because of this, processes with four-gluon vertices are not considered since its color structure is already accounted for by a proper treatment of the three-gluon vertices.
\end{enumerate}

These observations reduce significantly the number of cases we have to consider to show that the inductive step in our proof indeed works. Suppose that all the correlators needed to describe all processes with $k$ particles in the final state in the large-$N_c$ limit are dipoles and quadrupoles. Now consider a planar diagram for a process with $k+1$ particles in the final state, remove one gluon and consider the representation of the leading $N_c$ piece of the remaining diagram in terms of only fundamental Wilson lines. Since it is a planar diagram corresponding to a process with $k$ particles in the final state its representation in terms of only fundamental Wilson lines is expressed in terms of dipoles and quadrupoles only. Now reattach the gluon removed in the previous step, there are two possibilities: either both of its legs are attached to the same fermion loop, or, each leg is attached to a different fermion loop. In the first case, one can see from the examples in Section \ref{sec:3part} that the insertion of the new gluon splits the corresponding fermion loop, either dipole or quadrupole, creating a new quadrupole (when the new dipole interacts both in the amplitude and the conjugate amplitude, as was pointed out any other case would lead to simpler correlators). The second case can be easily seen to be suppressed by a factor of $1/N_c^2$ with respect to the first case. The two terms one obtains after applying the Fierz identity to a gluon joining two separate fermion loop have one power of $N_c$ less as compared to the diagram without the extra gluon, one of them has one loop less while the other has an explicit factor of $1/N_c$. This is to be compared to the first case where one additional fermion loop is created, and therefore one extra power of $N_c$ is present.

We can therefore conclude that for any multiple-jet graphs, under the framework of saturation physics, the dipole and quadrupole are the only objects appear in a physical multiple-jet production process in the large $N_c$ limit. One should note that so far, our proof does not apply to multi-particle production processes with large rapidity intervals or gaps between the measured particles (all jets are produced in the same rapidity region). For such situations, one needs to consider higher orders diagrams (and at least re-sum those that contain a logarithmic enhancement). In the following section, 
we explain that our proof holds to all orders in $\alpha_s$, which encompasses those situations where particles are emitted with large rapidity differences.

\subsection{Higher order $\alpha_s$ corrections}
\begin{figure}[tbp]
\begin{center}
\includegraphics[width=8.4cm]{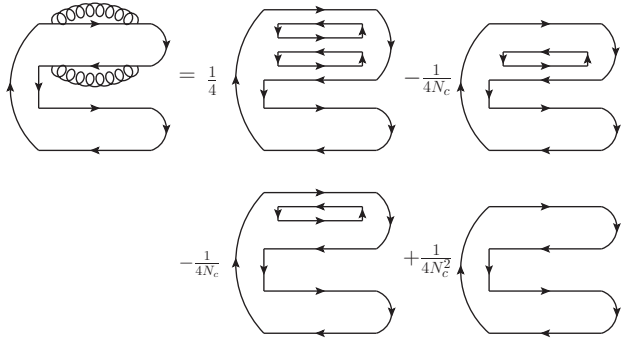}
\end{center}
\caption[*]{Higher order $\alpha_s$ contribution to the $q\bar q$ dijet}
\label{virtual}
\end{figure}
As briefly mentioned earlier, within the framework of the dilute-dense factorization, the next-to-leading order $\alpha_s$ corrections  of the single-inclusive production cross sections~\cite{Chirilli:2011km} in the large $N_c$ limit does not involve higher point functions. Here we would like to generalize this conclusion up to all order. There are two classes of graphs at higher order, namely, the virtual graphs and real graphs. In terms of color structure in the coordinate space, these two classes of graphs are actually similar. They both introduce an additional gluon at a new coordinate. The only difference is that the additional gluon in the real graphs goes through cut and is produced in the final state. For the real graphs, we can follow the above discussion on the multiple-jet productions, and integrate over arbitrary number of final state jets, and arrive at the conclusion that all real graphs in the large $N_c$ limit can only involve dipoles and quadrupoles. For the virtual graphs, we can easily see that the large $N_c$ limit requires the additional gluon within a dipole or a quadrupole, thus higher order virtual graphs can only generate more dipoles. 
Let us take the DIS dijet as an example, where we add one-gluon loop in both of the amplitude and complex conjugate amplitude as shown in Fig.~\ref{virtual}. Using the Fierz identity, it is obvious that the addition of the virtual gluons in the dijet processes does not introduce higher point functions in the large $N_c$ limit.

We have shown that, to all orders in $\alpha_s$, all multi-particle production processes in dilute-dense collisions are expressible in terms of dipoles and quadrupoles only in the large-$N_c$ limit. In particular, this includes the case where particles are emitted with a large rapidity difference, along with un-tagged emissions or gaps in between. We would like to point out that, in the case of strongly-ordered gluon emissions, this result was first obtained in Ref.~\cite{Kovner:2006wr}. Our derivation extends the result to the general case of non-eikonal, gluon, quark or antiquark emissions.

\section{Conclusion}

In conclusion, we find that in the large $N_c$ limit, all multi-particle production processes in p+A-type collisions up to all orders in $\alpha_s$ can be described in terms of only dipoles and quadrupoles under the CGC framework, excluding cases with large rapidity intervals or gaps between the measured particles. This conclusion can very possibly lead us to an effective $k_t$ factorization at small-$x$ for multiple-jet production processes in high energy scatterings within a dilute-dense system. This effective $k_t$ factorization involves two fundamental objects, namely, the dipole and quadrupole, and it can only work in the large $N_c$ limit. Only in the large $N_c$ limit, can one get rid of all the higher point functions which are presumably new objects. Provided we understand both dipoles and quadrupoles well, we will be able to predict any multiple-jet production processes up to corrections of order $\frac{1}{N_c^2}$ by using this effective $k_t$ factorization.

\begin{acknowledgments}
We thank F. Gelis, A. Kovner, A. H. Mueller, R. Venugopalan and F. Yuan
for discussions and comments. This work was supported in part by
the U.S. Department of Energy under the contracts DOE OJI grant No. DE - SC0002145 and Polish NCN grant DEC-2011/01/B/ST2/03915.
FD's research is supported by the European Research Council under the Advanced Investigator Grant ERC-AD-267258.
CM's research is supported by the European Research Council grant HotLHC ERC-2011-StG-279579.
AMS is supported by the Sloan Foundation.

\end{acknowledgments}

\appendix

\section{Large-$N_c$ evaluation of correlators in the McLerran-Venugopalan model}

In this Appendix we show how one can compute general $n$-point correlators under the framework of the McLerran-Venugopalan model\cite{McLerran:1993ni} in the large-$N_c$ limit. The formalism we will employ is the one introduced in Refs. \cite{HiroFujii,Blaizot:2004wv} which has been successfully used to compute the full finite-$N_c$ expressions for several correlators also in \cite{Dominguez:2008aa,Dominguez:2011wm,Kovchegov:2008mk,Marquet:2010cf}.

The general strategy consists on expanding the Wilson lines and then take advantage of the fact that the only non-trivial correlation is the average of two gauge fields by using Wick's theorem. The elementary correlator of two fields takes the form
\begin{equation}
g_S^2\langle A_c^-(z^+,x)A_d^-(z^{\prime+},y)\rangle_{x_g}=\delta_{cd}\delta(z^+-z^{\prime+})\mu_{x_g}^2(z^+)L_{xy},
\end{equation}
where $L_{xy}$ can be written in terms of a two-dimensional massless propagator (see Ref. \cite{Dominguez:2011wm}). In order to be able to resum these two-point contractions it is necessary to pay close attention to the color algebra. Only overall singlet states need to be considered, one can therefore calculate the matrix indicating the possible transitions between such states and then diagonalize it.

This was done explicitly for correlators involving four fundamental Wilson lines (two quark-antiquark pairs) in \cite{Blaizot:2004wv,Dominguez:2008aa,Dominguez:2011wm} where only two overall singlet states are available and the corresponding transition matrix takes the following form
\begin{equation}
M=\left(\begin{matrix}C_F(L_{x_1x_2}+L_{x'_2x'_1})+\frac{1}{2N_c}F(x_1,x_2;x'_2,x'_1) & -\frac{1}{2}F(x_1,x'_1;x'_2,x_2)\\ -\frac{1}{2}F(x_1,x_2;x'_2,x'_1) & C_F(L_{x_1x'_1}+L_{x'_2x_2})+\frac{1}{2N_c}F(x_1,x'_1;x'_2,x_2)\end{matrix}\right),
\end{equation}
with $F(x_1,x_2;x'_2,x'_1)=L_{x_1x'_2}-L_{x_1x'_1}+L_{x_2x'_1}-L_{x_2x'_2}$.

If one naively takes the large-$N_c$ limit the matrix becomes diagonal, which is consistent with the fact that each color transition is suppressed by a factor of $1/N_c^2$. For correlators where the singlet structure is not the same at both ends of the longitudinal extent it is not appropriate to take the large-$N_c$ limit at the level of the transition matrix, since different singlet states receive different weights in the final calculation. For example, for the quadrupole calculation in \cite{Dominguez:2011wm} one can see that the relevant combination of matrix elements for the $n$th order term of the expansion takes the form $(M^n)_{11}+N_c(M^n)_{21}$, where the non-diagonal component appears multiplied by a factor of $N_c$.

One can see from the explicit expression for the matrix $M$ that $M_{11}$ and $M_{22}$ have one additional factor of $N_c$ when compared to $M_{12}$ and $M_{21}$. When taking the leading $N_c$ term of the elements of $M^n$ it is easy to see that $(M^n)_{11}=M_{11}^n$ and $(M^n)_{22}=M_{22}^n$ while $(M^n)_{21}$ includes only terms with one factor of $M_{21}$ and $n-1$ factors of $M_{11}$ or $M_{22}$ which is the same as taking $M_{12}=0$ right from the beginning ($M_{11}$ and $M_{22}$ can be replaced by their large-$N_c$ versions also). The fact that the non-diagonal element enters only once signals that there was only one color transition.

One can easily see that
\begin{align}
(M^n)_{21}&=M_{21}\sum_{k=0}^{n-1}M_{11}^kM_{22}^{n-k-1},\nonumber\\
&=M_{21}\frac{M_{11}^n-M_{22}^n}{M_{11}-M_{22}}.
\end{align}

One must perform an ordered integral of the longitudinal coordinates and then sum over $n$. Since in a MV-like Gaussian model the longitudinal dependence of the correlations factors out, the ordered integral is equal to $\frac{1}{n!}$ times the full integral. From there it is easy to see that the $n$th powers appearing above become exponentials and the large-$N_c$ formula for the quadrupole is recovered.

This way of taking the large-$N_c$ limit at the matrix level can be generalized to more complicated correlators with similar results. One can find an ordering of the singlet states such that all the necessary information to find the large-$N_c$ version of the correlator is in a lower diagonal matrix organized in blocks in which going away from the diagonal lowers the power of $N_c$ of the corresponding matrix element. Here we describe this procedure for the case of the six-point correlator of the form $\frac{1}{N_c}\langle\text{Tr}(U_1U^\dagger_2U_3U^\dagger_4U_5U^\dagger_6)\rangle$.

\begin{figure}[tbp]
\begin{center}
\includegraphics[width=\textwidth]{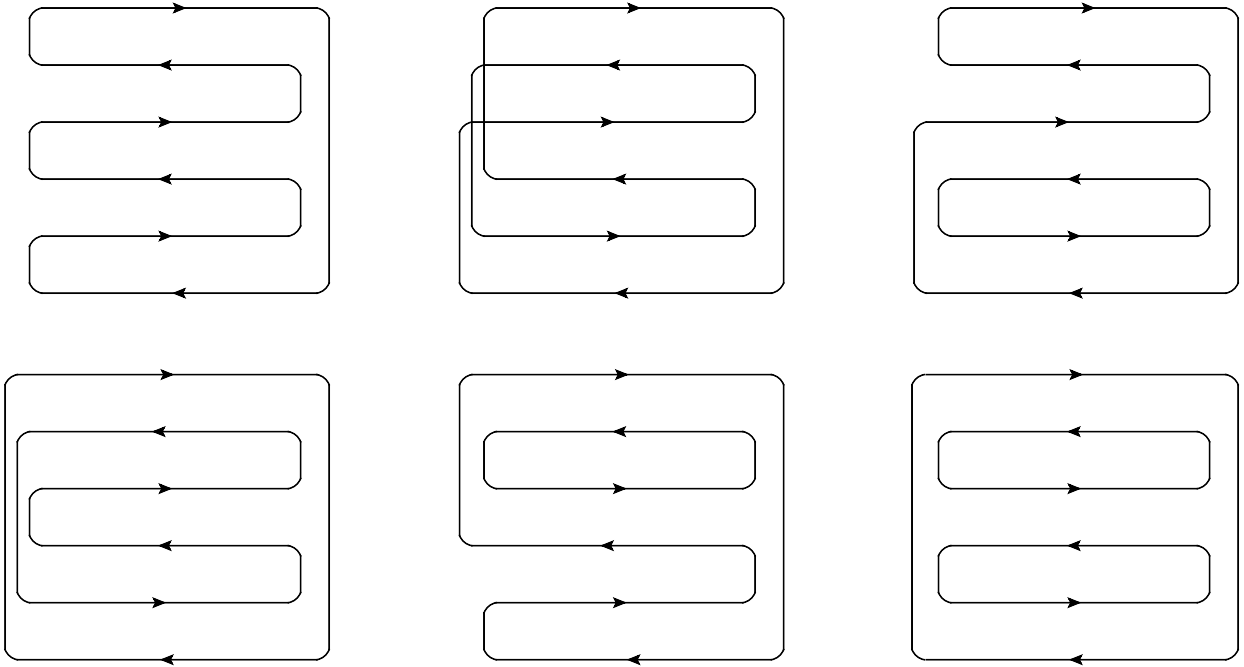}
\end{center}
\caption[*]{Graphical representation of the six topologies involved in the calculation. The power of $N_c$ associated to each configuration is equal to the number of fermion loops.}
\label{singlets}
\end{figure}

For such a system of three quarks and three antiquarks there are 6 singlet states available corresponding to the 6 ways of pairing quarks with antiquarks. The proper way to organize the states in the matrix representation is according to the power of $N_c$ of the overlap with the final singlet state. In terms of the graphical representation introduced in \cite{Blaizot:2004wv}, the six singlet states correspond, in that particular order, to the topologies shown in Fig. \ref{singlets} where the number of fermion loops gives the power of $N_c$ associated to the overlap.

In the large-$N_c$ limit only transitions to states corresponding to a higher power of $N_c$ are allowed, which, thanks to the specific order chosen for the singlet states, means that one can divide the corresponding transition matrix into blocks and all the blocks above the diagonal can be neglected. The corresponding transition matrix $M$ takes then the following form
\begin{equation}
M=\left(\begin{matrix}M_1 & 0 & 0\\ M_4 &M_2 & 0\\ 0 & M_5 & M_3\end{matrix}\right),
\end{equation}
where the matrices along the diagonal are diagonal, and the off-diagonal matrices have one power of $N_c$ less than the ones along the diagonal. It is not difficult to calculate the $n$th power of this matrix
\begin{equation}
M^n=\left(\begin{matrix}M_1^n & 0 & 0\\ \sum_{i=0}^{n-1}M_2^iM_4M_1^{n-i-1} & M_2^n & 0\\ \sum_{i=0}^{n-2}\sum_{j=0}^{n-i-2}M_3^iM_5M_2^jM_4M_1^{n-i-j-2} & \sum_{i=0}^{n-1}M_3^iM_5M_2^{n-i-1} & M_3^n\end{matrix}\right).
\end{equation}
One can easily see that the relevant matrix elements for the evaluation of the desired correlator are of the form
\begin{equation}
\left(\begin{matrix}1&1&N_c&N_c&N_c&N_c^2\end{matrix}\right)M^n\left(\begin{matrix}1\\ 0\\ 0\\ 0\\ 0\\ 0\end{matrix}\right).\label{melmsext}
\end{equation}
This singles out the first column of $M^n$. Since $M_1$ and $M_2$ are diagonal matrices it is easy to see that the elements of $M$ in the second column never enter the expressions for the elements of $M^n$ in the first column. Following this observation, we drop completely all contributions from the second column (and second row) which correspond to the second topology proportional to $N_c$.

As a 5x5 matrix, $M$ has the form
\begin{equation}
M=\begin{pmatrix}m_{11} & 0 & 0 & 0 & 0\\ m_{21} & m_{22} & 0 & 0 & 0\\ m_{31} & 0 & m_{33} & 0 &0\\ m_{41} & 0 & 0 & m_{44} & 0\\ 0 & m_{52} & m_{53} & m_{54} & m_{55}\end{pmatrix}.
\end{equation}
The first column of its $n$th power is
\begin{equation}
\begin{pmatrix} m_{11}^n\\ m_{21}\sum_{i=0}^{n-1}m_{11}^im_{22}^{n-i-1}\\ m_{31}\sum_{i=0}^{n-1}m_{11}^im_{33}^{n-i-1}\\ m_{41}\sum_{i=0}^{n-1}m_{11}^im_{44}^{n-i-1}\\ \sum_{k=2}^4m_{5k}m_{k1}\left[\sum_{i=0}^{n-2}\sum_{j=0}^{n-i-2}m_{11}^im_{kk}^jm_{55}^{n-i-j-2}\right]\end{pmatrix},
\end{equation}
which can be rewritten as
\begin{equation}
\begin{pmatrix} m_{11}^n\\ \frac{m_{21}}{m_{11}-m_{22}}\left[m_{11}^n-m_{22}^n\right]\\ \frac{m_{31}}{m_{11}-m_{33}}\left[m_{11}^n-m_{33}^n\right]\\ \frac{m_{41}}{m_{11}-m_{44}}\left[m_{11}^n-m_{44}^n\right]\\ \sum_{k=2}^4m_{5k}m_{k1}\left[\frac{m_{11}^n}{(m_{11}-m_{kk})(m_{11}-m_{55})}+\frac{m_{kk}^n}{(m_{kk}-m_{11})(m_{kk}-m_{55})}+\frac{m_{55}^n}{(m_{55}-m_{kk})(m_{55}-m_{11})}\right]\end{pmatrix}.\label{nth1col}
\end{equation}
Now, as in the previous case, these expressions appear in the final result summed over $n$ with a factor of $\frac{1}{n!}$ due to the ordering in the longitudinal coordinate. Therefore the $n$th powers become exponentials.

One can easily calculate explicitly the relevant leading-$N_c$ piece of the transition matrix for the case of interest. This reduced version of the transition matrix then takes the form
{\scriptsize
\begin{equation}
M=\frac{1}{2}\begin{pmatrix}N_c(L_{12}+L_{34}+L_{56})& 0 & 0 & 0 & 0\\ F_{1243} & N_c(L_{14}+L_{32}+L_{56})& 0 & 0 & 0\\ F_{1265} & 0 & N_c(L_{16}+L_{34}+L_{52})& 0 & 0\\ F_{3465} & 0 & 0 & N_c(L_{12}+L_{36}+L_{54})& 0\\ 0 & F_{1465} & F_{2534} & F_{1263} & N_c(L_{16}+L_{32}+L_{54})\end{pmatrix}
\end{equation}
}
where $F_{ijkl}=L_{ik}-L_{jk}+L_{jl}-L_{il}$. Plugging these matrix elements back into the above equations, transforming the $n$th powers into exponentials, and including the tadpole contributions one gets
\begin{eqnarray}
&&\frac{1}{N_c}\left\langle\text{Tr}\left[ U_{1}U_{2}^{\dagger }U_{3}U_{4}^{\dagger
}U_{5}U_{6}^{\dagger }\right]\right\rangle\nonumber \\
&=&e^{-\Gamma_{12}-\Gamma_{34}-\Gamma_{56}}-\frac{F_{1234}}{F_{1324}}\left[e^{-\Gamma_{12}-\Gamma_{34}}-e^{-\Gamma_{14}-\Gamma_{32}}\right]e^{-\Gamma_{56}}\nonumber\\
&&-\frac{F_{1256}}{F_{1526}}\left[e^{-\Gamma_{12}-\Gamma_{56}}-e^{-\Gamma_{16}-\Gamma_{52}}\right]e^{-\Gamma_{34}}-\frac{F_{3456}}{F_{3546}}\left[e^{-\Gamma_{34}-\Gamma_{56}}-e^{-\Gamma_{36}-\Gamma_{54}}\right]e^{-\Gamma_{12}}\nonumber\\
&&+F_{1234}F_{1456}\left[\frac{e^{-\Gamma_{12}-\Gamma_{34}-\Gamma_{56}}}{F_{1324}G}-\frac{e^{-\Gamma_{14}-\Gamma_{32}-\Gamma_{56}}}{F_{1324}F_{1546}}+\frac{e^{-\Gamma_{16}-\Gamma_{32}-\Gamma_{54}}}{F_{1546}G}\right]\nonumber \\
&&+F_{1256}F_{2543}\left[\frac{e^{-\Gamma_{12}-\Gamma_{34}-\Gamma_{56}}}{F_{1526}G}-\frac{e^{-\Gamma_{16}-\Gamma_{34}-\Gamma_{52}}}{F_{1526}F_{2453}}+\frac{e^{-\Gamma_{16}-\Gamma_{32}-\Gamma_{54}}}{F_{2453}G}\right]\nonumber \\
&&+F_{3456}F_{1236}\left[\frac{e^{-\Gamma_{12}-\Gamma_{34}-\Gamma_{56}}}{F_{3546}G}-\frac{e^{-\Gamma_{12}-\Gamma_{36}-\Gamma_{54}}}{F_{3546}F_{1326}}+\frac{e^{-\Gamma_{16}-\Gamma_{32}-\Gamma_{54}}}{F_{1326}G}\right],
\end{eqnarray}
where $\Gamma_{ij}=\mu^2(L_{ii}+L_{jj}-2L_{ij})$ and $G=L_{12}+L_{34}+L_{56}-L_{16}-L_{32}-L_{54}$.

In this expression one can easily recognize three different type of contributions depending on the number of transition between the singlet states. The first term clearly comes from the first row in (\ref{nth1col}) and corresponds to the totally elastic part with no color transitions, the following three terms come from the second, third, and fourth rows of (\ref{nth1col}) and correspond to terms with only one transition, while the rest of the terms come from the last row of (\ref{nth1col}) and are associated with terms which have two transitions between singlet states.

\end{document}